# Mage: Online Interference-Aware Scheduling in Multi-Scale Heterogeneous Systems

Francisco Romero
Electrical Engineering Department
Stanford University
faromero@stanford.edu

Christina Delimitrou
Electrical and Computer Engineering
Cornell University
delimitrou@cornell.edu

## ABSTRACT

Heterogeneity has grown in popularity both at the core and server level as a way to improve both performance and energy efficiency. However, despite these benefits, scheduling applications in heterogeneous machines remains challenging. Additionally, when these heterogeneous resources accommodate multiple applications to increase utilization, resources are prone to contention, destructive interference, and unpredictable performance. Existing solutions examine heterogeneity either across or within a server, leading to missed performance and efficiency opportunities.

We present Mage, a practical interference-aware runtime that optimizes performance and efficiency in systems with intra- and inter-server heterogeneity. Mage leverages fast and online data mining to quickly explore the space of application placements, and determine the one that minimizes destructive interference between co-resident applications. Mage continuously monitors the performance of active applications, and, upon detecting QoS violations, it determines whether alternative placements would prove more beneficial, taking into account any overheads from migration. Across 350 application mixes on a heterogeneous CMP, Mage improves performance by 38% and up to 2x compared to a greedy scheduler. Across 160 mixes on a heterogeneous cluster, Mage improves performance by 30% on average and up to 52% over the greedy scheduler, and by 11% over the combination of Paragon [15] for inter- and intra-server heterogeneity.

## 1 INTRODUCTION

The challenges that emerge from the end of Dennard scaling [22] have motivated an extensive line of work on architecture heterogeneity. Incorporating heterogeneity, whether at an individual resource (core, memory), or server platform granularity, allows the system to better match applications to the underlying hardware, given each heterogeneous resource's capabilities and limitations [11, 34, 41]. Apart from core and memory heterogeneity, platform heterogeneity has become increasingly the norm in large-scale cloud infrastructures [5, 15–17, 19, 21, 55]. As servers get progressively replaced and upgraded during a datacenter's lifetime, the system can end up with several tens of different platform generations and configurations. Heterogeneity across and within servers presents an interesting challenge, as co-scheduled applications experience interference in shared resources. Cloud platforms often employ multi-tenancy to increase system utilization. Unfortunately this also leads to unpredictable performance due to resource interference. While interference is present in homogeneous architectures as well, it becomes more challenging in the presence of heterogeneity, as the scheduler must also account for the impact of heterogeneity on resource contention [18, 41].

Both heterogeneity and interference contribute to performance unpredictability, which results in violations of the strict quality of service (QoS) requirements most cloud applications have. To eliminate unpredictable performance, one must address two challenges. First, the scheduler must determine the performance an application will achieve on each of the different heterogeneous resources or platforms, in the presence of interference from co-scheduled applications. Initial placement needs to happen fast to avoid high scheduling overheads at admission control. Second, the scheduler must revisit its initial placement decisions to adapt to application churn, changes in application behavior, and to correct poor initial placements. Rescheduling must incur minimal performance overheads and, when migration is needed, it needs to distinguish between stateless and stateful applications.

Related work has proposed several approaches to tackle heterogeneity and interference. For example, BubbleFlux [41] determines the sensitivity of applications to memory pressure in order to co-schedule high-priority services with appropriate best-effort workloads. Similarly, Paragon [15] uses fast classification techniques to determine the most suitable server platform for a given application, and its sensitivity to interference in shared resources. While these approaches accurately capture an application's resource preferences, they focus on platform-level heterogeneity, which is a relatively small design space (a few tens of platforms). When additionally considering heterogeneity at the granularity of individual resources, a lot of previously-proposed mechanisms do not scale without substantially increased runtime overheads.

In this work we improve resource efficiency while preserving QoS through a tiered scheduling approach that takes advantage of heterogeneity both within and across servers. Specifically, we propose Mage, a practical runtime system that leverages a set of scalable and online machine learning techniques based on latent factor models, namely stochastic gradient descent (SGD), to determine the performance of each scheduled application under different application-to-resource mappings, in the presence of heterogeneity and multi-tenancy. The techniques in Mage are transparent to the user, lightweight, and scale linearly with the number of applications.

Mage is a hierarchical scheduler, first determining the most suitable server platform across the datacenter, and then allocating appropriate heterogeneous resources within the server, additionally trading off inter- for intra-server heterogeneity when needed. Once a workload is scheduled, Mage monitors its performance and reacts to discrepancies between estimated and measured performance within a few milliseconds. When the performance of an active application deviates from its expected value, Mage evaluates alternative application placements and determines whether the performance improvement from re-scheduling one or more workloads outweighs the corresponding performance penalties.

The main contributions of Mage are the following:

- Bridge CMP- and cluster-level heterogeneous scheduling via a tiered approach that outperforms the sum of systems addressing these problems independently.
- Introduce a *staged*, parallel SGD to determine the resource requirements of a new application, which greatly reduces complexity, allowing the scheduler to scale to hundreds of applications, and to explore the entire space of application placements. This removes the need for empirically decomposing scheduling to smaller, independently-solved sub-problems, which keep overheads low, but result in suboptimal performance, and require substantial allocation adjustments at runtime.
- Enable fast and lightweight re-scheduling, in the event that QoS violations are observed. Because Mage considers heterogeneity at the server and cluster level jointly, it can re-allocate resources in a less invasive way than systems accounting for these two problems separately.

We have evaluated Mage using both simulation and experiments on a 40-server heterogeneous cluster, with latency-critical services, and batch workloads. We compare Mage against five schedulers: (i) a greedy scheduler that allocates the largest available server (or core) first, (ii) a power-aware scheduler that allocates the lowest-power platform first, (iii) a static version of Mage, *Mage-Static*, where decisions are made at admission and not revisited thereafter, (iv) PIE [52], a heterogeneity-aware CMP scheduler, and (v) Paragon [15], a heterogeneity-aware cluster scheduler.

We evaluate three execution scenarios. First, we evaluate Mage in a heterogeneous 16-core CMP, through simulation.

Across 350 diverse application mixes, Mage improves performance by 38% compared to the greedy scheduler, by 61% compared to the power-efficient scheduler, by 33% compared to PIE, and by 13% compared to Paragon. It also outperforms Mage-Static by 22% of average, as the latter cannot address changes in application behavior at runtime, such as unexpected spikes in user load, or incorrect initial placements. Second, we deploy Mage in a real heterogeneous 40-server cluster. Mage again improves performance by 30% compared to the greedy scheduler, and by 51% compared to the power-efficient scheduler. Third, we demonstrate the tiered behavior of Mage by introducing core-level heterogeneity in the real cluster through power management. In this case, Mage outperforms the greedy scheduler by 45%, and the power-efficient scheduler by 56%. We also compare Mage against a combination of Paragon for server-level heterogeneity and PIE for core-level heterogeneity, and show 19% improvement with Mage. Finally, we use Paragon both at the cluster and server level, and show that having a unified scheduling framework like Mage outperforms schedulers that independently handle inter- and intra-server heterogeneity, even if each scheduler is heterogeneity- and interference-aware. Mage outperforms Paragon+Paragon by 11% on average because it has a global view of resource availability, and can trade off core- (frequency) for server-level heterogeneity whenever necessary.

Finally, we show that the more heterogeneous a system is, the higher the benefit from Mage. For instance, in a cluster with 2 server types, the performance benefit of Mage versus a greedy scheduler is 15% on average. When the same cluster has 10 server types the benefit increases to 75%. As systems become increasingly heterogeneous, runtimes like Mage can ensure that the added heterogeneity does not come at a performance and/or efficiency loss.

## 2 RELATED WORK

**Heterogeneity-aware datacenter scheduling:** Datacenters are becoming increasingly heterogeneous at the server level. With the datacenter building provisioned for a 15-year lifetime, and servers progressively replaced and upgraded over that period, it is not uncommon for cloud systems to consist of a few tens of server generations and configurations. This heterogeneity can have a significant impact on application performance, especially for interactive, latency-critical applications [15, 18, 44, 55]. The most closely related work to Mage is Paragon, which leverages practical classification techniques to determine which server platform is best suited for an incoming, unknown application [15]. Whare-map also quantifies the impact of server heterogeneity on performance and cost for a set of Google production workloads [41]. Similarly, Nathuji et al. leverage server heterogeneity to improve datacenter power efficiency [44]. While these systems can correctly identify the platforms that optimize the performance and/or energy efficiency of cloud applications at the granularity of servers, they are not lightweight enough to make decisions at core granularity. As cloud servers start incorporating heterogeneity in compute [30, 47], memory, and storage, it is essential to have scalable schedulers that account for heterogeneity and make high quality decisions online.

**Contention-aware datacenter scheduling:** Sharing system resources to increase utilization results in interference, which translates to performance degradation [15, 18, 41, 46], and in some cases security vulnerabilities [20]. Several recent systems aim to minimize destructive interference by disallowing colocation of jobs that contend in the same resources [15, 18, 41, 42, 45, 46], or by partitioning resources to improve isolation [20, 31, 32, 39]. For example, BubbleFlux determines how the sensitivity of applications to pressure in memory resources evolves over time, and prevents multiple memory-intensive applications from sharing the same platform. Similarly, DeepDive identifies the interference VMs sharing resources experience, and manages it transparently to the user [46]. In the same spirit, Nathuji et al. [45] develop Q-Clouds, a QoS-aware control framework that dynamically tunes resources in virtualized clouds to mitigate resource interference between contending VMs. On the isolation front, Lo et al. [39] study the sensitivity of Google applications to different sources of interference. They then combine hardware and software isolation techniques, including containers, cache partitioning, network bandwidth partitioning, and power management [38] to preserve QoS for the latency-critical, interactive application, when they share resources with batch, low-priority workloads.

**Heterogeneous CMP scheduling:** The end of process scaling has made compute and memory heterogeneity highly relevant to modern multicores [11, 23, 25, 27, 40, 47, 51, 52]. To manage this fine-grained heterogeneity, several recent schedulers and runtime systems account for the impact of heterogeneity on application performance at the hypervisor [24], OS [23, 25, 51], or hardware-level [11, 40, 52]. Shelepov et al. [51], for example, present a heterogeneity-aware scheduler that achieves good performance and fairness, and is simple and scalable, while Craeynest et al. [52] use performance statistics to find application placements that offer the highest performance in CMPs with big and small cores. Scheduling datacenter applications shares a lot of commonalities with multi-core scheduling, with the added challenge that cloud services care about tail latency as opposed to average throughput. This makes interference equally critical to heterogeneity.

# 3 MAGE DESIGN
## 3.1 Overview

The key requirement for interference-aware scheduling in heterogeneous systems is determining the impact of heterogeneity and interference on application performance in an accurate and fast manner. Previous work has either tackled heterogeneity and/or interference at platform granularity [15, 41, 55], or has exclusively managed core heterogeneity within a server via per-core performance models and OS runtimes [11, 23, 25, 40, 51, 52]. Unfortunately managing heterogeneity at the granularity of entire servers poses more relaxed constraints on scheduling overheads than when heterogeneity is at the granularity of individual cores. Additionally, accounting for heterogeneity and interference separately ignores the performance impact the former has on the latter. This means that either performance will be suboptimal, or several scheduling decisions will have to be revisited at runtime, incurring penalties from migration. Finally, while per-core performance models can accurately predict the difference in performance between high-end and low-power cores [40, 52], they do so with applications running in isolation, hence they do not account for the interference between co-scheduled applications. Additionally, because they rely on detailed analytical models, they can be computationally expensive when applied across a cloud infrastructure. As systems become increasingly heterogeneous, we need accurate, scalable, and lightweight runtimes that optimize for both performance and efficiency.

Mage is a runtime that accounts for heterogeneity and interference jointly, at the server and datacenter level. It operates as a *tiered scheduler*, providing a unified framework for managing heterogeneity within and across server platforms. Mage first determines the appropriate platform for an application in a heterogeneous cloud, and then determines the most suitable among different heterogeneous resources of a single server. To simplify our discussion, we will focus on heterogeneous cores for now, and expand Mage to other resources in Section 6.2. Instead of employing performance models, Mage follows an architecture-agnostic, *data-driven approach*. Specifically, it leverages a set of practical machine learning techniques that rely on a minimal amount of profiling information to infer an application's performance on any of several heterogeneous resources *in the presence* of interference. Mage introduces a *staged* latent factor model using Stochastic Gradient Descent (SGD) [8, 33, 53] to infer the impact of interference when new unknown applications run on heterogeneous resources.

Figure 1 shows an overview of the system. Once a new application arrives, it is profiled for a few seconds to provide SGD with a sparse signal of the application's resource requirements. SGD then uses this signal to determine how the application will behave on any of the available heterogeneous resources, given how previous, similar applications behaved on them. Subsequently, Mage uses this information to select the application-to-core mapping that maximizes performance across co-scheduled applications. Once applications are scheduled, Mage continuously monitors their behavior. If one or more applications experience suboptimal performance, they are reprofiled, reclassified, and potentially rescheduled.

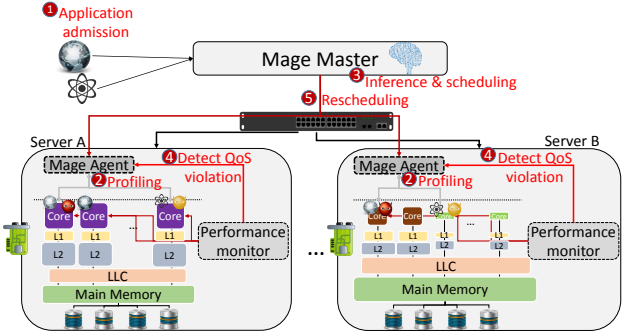

**Figure 1: Overview of the Mage runtime.**

In the following sections we discuss the ML techniques, overheads and scalability of Mage, and present a validation that ensures that the inferred performance reflects the application's behavior once scheduled.

*3.1.1 Machine Learning Background*

Stochastic Gradient Descent (SGD) is a popular latent factor model in online ML systems [6, 8, 33, 37, 48, 49, 53, 56]. It is primarily used as a preprocessing step to matrix factorization, in recommender systems and online classifiers [6], and more recently to train weights and biases in neural networks. SGD has become the training model of choice primarily because of its efficiency in the presence of massive datasets. Large datasets have been a major roadblock for previously-used techniques, such as the Interior-Point, or Newton Method [28], which rely on basic linear algebra routines like Cholesky, LU, or DGEMM, and become computationally prohibitive when datasets increase. The input to SGD is an m×d matrix $A$, called the utility matrix. Each row (or sample) $r$ in $A$ corresponds to an instance of the training set containing $d$ features (the dimensionality of the dataset), and each column corresponds to different items to be recommended. The goal of SGD is to find a $d$-dimensional vector $w$, which minimizes objective function, $f$, also known as loss function. In Mage, the loss function is the error between inferred performance, and measured performance on a real heterogeneous platform.

SGD is an iterative process. In every iteration SGD computes the gradient of the objective function with respect to each entry in the utility matrix, and makes an update to the model in the negative direction of the gradient. The initial utility matrix $A$ is heavily sparse. Before SGD can iterate over each element of $A$, we need to provide it with an approximation of the dense utility matrix $R$ using PQ-reconstruction [8, 53], where $R \approx Q \cdot P^T$. A popular approach to obtain $P$ and $Q$ is through matrix factorization, e.g., via a technique like Singular Value Decomposition (SVD) [6], under which $Q_{m \times r} = U$, and $P^T_{r \times d} = \Sigma \cdot V^T$. $U_{m \times r}$ is the matrix of left singular vectors, $\Sigma_{r \times r}$ is the diagonal matrix of singular values, and $V_{d \times r}$ is the matrix of right singular vectors. Once we obtain the initial $R$, SGD progressively improves its per-element estimations:

$\forall r_{ui}$, where $r_{ui}$ an element of the reconstructed matrix $R$
$\varepsilon_{ui} = r_{ui} - q_i \cdot p_u^T$
$q_i \leftarrow q_i + \eta(2 \cdot \varepsilon_{ui} p_u - \lambda q_i)$
$p_u \leftarrow p_u + \eta(2 \cdot \varepsilon_{ui} q_i - \lambda p_u)$
until $|\varepsilon|_{L_2} = \sqrt{\sum_{u,i} |\varepsilon_{ui}|^2}$ becomes marginal.

$\eta$ is the learning rate, and $\lambda$ is the regularization parameter. The learning rate determines how quickly the values adapt between iterations, thus affects the speed of convergence, and the regularization parameter avoids overfitting to a specific dataset. The learning rate is selected to be small enough for SGD to achieve linear convergence. It is set at $(\lambda \cdot k^{-1})$ where $k$ the current iteration, which can be proven to be optimal, since the Hessian of the cost function at the optimum is strictly positive [9]. The regularization parameter is selected empirically for a given dataset. The complexity of SGD is $O(k \cdot m \cdot p)$, where $k$ the number of iterations, $m$ the number of rows of the utility matrix, and $p$ the average number of non-zero entries per row (*density degree*).

**Parallel SGD:** A major advantage of SGD is its significantly lower complexity compared to prior techniques. Nevertheless, scheduling applications on heterogeneous multicores poses strict constraints on scheduling overheads, both during the initial placement, and any runtime adjustments. Under these constraints, SGD can still add considerable overheads, especially for large datasets. To further improve the scalability of Mage, we implement *parallel SGD* [37, 49], which leverages shared memory to achieve near-linear speedup with the number of processors. When running Mage on a dedicated server, parallel SGD reduces scheduling overheads by at least an order of magnitude.

## 3.2 Heterogeneous Scheduling with SGD

The input to Mage is profiling information of incoming applications with co-scheduled applications, either targeted contentious kernels, or other cloud applications. The output is performance across application placements. Both Paragon [15] and Quasar [18] demonstrated that SGD can be used to classify unknown applications with respect to different server configurations and sources of interference. Mage improves on these findings in two ways. First, to improve inference scalability, Mage uses parallel SGD. Second, to further reduce the overheads of scheduling, Mage introduces a *staged* approach in parallel SGD with three phases, one used for training and two for inference, as illustrated in Figure 2.

**SGD 1 (*offline training*):** First, Mage obtains the interference profile of a new application. When an application arrives to the system, it is profiled on any of the available platforms for a few seconds alone and with two contentious kernels. Each contentious kernel targets a specific shared resource (CPU, cache hierarchy, memory capacity and bandwidth, network bandwidth, storage capacity and bandwidth) and introduces pressure of tunable intensity to that resource [14]. The profiling microbenchmarks and their corresponding intensity are selected at random. We use the same set of contentious kernels across all incoming applications, which serve as a common reference point that trains the scheduler to the differences in the characteristics of new services. Mage collects the application performance in MIPS and inserts it as a new row in utility matrix $A1_{m \times q}$ below, where $m$ is the number of incoming applications, and $q$ the number of contentious kernels multiplied by their intensity plateaus (10-100% in 10% increments with 100% saturating the entire shared resource), plus one for the run in isolation.

$$A1_{m \times q} = \begin{array}{c} \\ app_1 \\ app_2 \\ app_3 \\ ... \\ app_m \end{array} \begin{bmatrix} alone & uB1_{10} & ... & uB1_{100} & uB2_{10} & ... & uBn_{100} \\ a1_{1,1} & a1_{1,2} & ... & 0 & 0 & ... & a1_{1,q} \\ a1_{2,1} & 0 & ... & a1_{2,11} & a1_{2,12} & ... & 0 \\ a1_{3,1} & 0 & ... & a1_{3,11} & 0 & ... & 0 \\ ... & ... & ... & ... & ... & ... & ... \\ a1_{m,1} & 0 & ... & 0 & a1_{m,12} & ... & 0 \end{bmatrix}$$

SGD recovers the performance for the missing entries and provides the scheduler with information on the sensitivity of the application to different types of interference across platforms. This SGD step only needs to happen *once* for a given application regardless of other applications present in the system at each point in time.

**SGD 2 (*online testing – partial placements*):** Once Mage obtains the interference profile for each incoming workload, it randomly selects an available core in a server, and executes the new application for 1-2 seconds. As the application is running, Mage collects performance statistics for this, as well as any already active applications on the same platform and populates the corresponding columns in the utility matrix. The utility matrix $A2$ - seen below - concatenates $A1$ with the profiled application-to-core mappings as columns. For example, $\text{map}_{1234}$ below means that $app_1$ is scheduled on $core_1$, $app_2$ on $core_2$, $app_3$ on $core_3$, and $app_4$ on $core_4$ of the same server. The obtained performance statistics enable Mage to recover the missing entries for columns $w$. Unfortunately, this only includes a small subset of all possible application-to-core mappings. The remaining mappings correspond to all-zero columns and need to tackled separately to avoid having them immediately discarded by SVD. Randomly initializing all-zero columns to increase entropy is a frequently-used approach in ML systems. However, in our case it increased scheduling overheads substantially, as more iterations were needed for SGD to converge, and additionally resulted in higher estimation errors. Instead we break the online inference to an additional step, SGD3, described below.

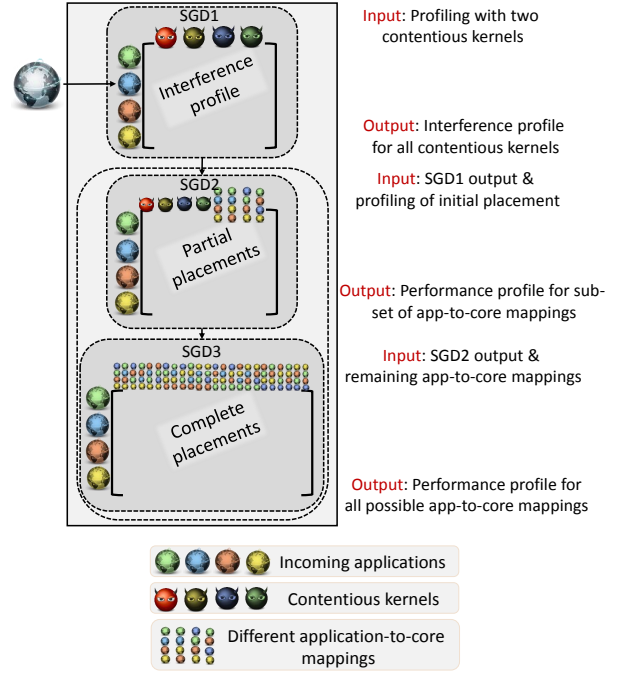

**Figure 2:** The staged SGD approach employed in Mage. SGD1 infers the sensitivity of incoming applications to resource contention, SGD2 determines per-application performance for a subset of placement strategies, and SGD3 infers performance for the remaining placements.

$$A2_{m \times (q+w)} = A1 \frown \begin{array}{c} app_1 \\ app_2 \\ app_3 \\ ... \\ app_m \end{array} \begin{bmatrix} \text{map}_{1234} & \text{map}_{1243} & ... & \text{map}_{4321} \\ a2_{1,1} & 0 & ... & 0 \\ 0 & a2_{2,2} & ... & 0 \\ 0 & 0 & ... & 0 \\ ... & ... & ... & ... \\ 0 & 0 & ... & a2_{m,(q+w)} \end{bmatrix}$$

**SGD 3 (*online testing – complete placements*):** A third and final SGD populates the previously all-zero columns $(n-w)$ of matrix $A3$ below. Because all other columns are now fully populated, randomly initializing the all-zero columns does not have the same negative impact on estimation error or complexity as before. The initialization range is $[min_{value}, max_{value}]$ of the existing matrix entries. Once this final SGD completes, Mage selects the placement (column) with the highest geometric mean, schedules the applications on the heterogeneous resources, and starts monitoring their performance. [1]

---
[1] A positive side-effect of breaking the online SGD into two steps is that the overall overhead is significantly lower, since the sum of the iterations for

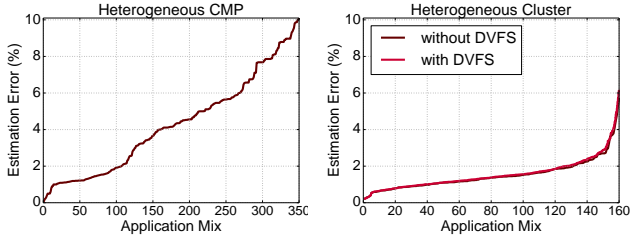

Figure 3: Geometric mean error per application mix, between the measured performance and the performance estimated by Mage for: (a) the simulated heterogeneous CMP, and (b) the heterogeneous cluster with and without power management.

$$A3_{m\times(q+n)} = A1 \frown \begin{array}{c} app_1 \\ app_2 \\ app_3 \\ ... \\ app_m \end{array} \begin{bmatrix} map_{1234} & map_{1243} & ... & map_{4321} \\ a2_{1,1} & a2_{1,2} & ... & a2_{1,(q+n)} \\ a2_{2,1} & a2_{2,2} & ... & a2_{2,(q+n)} \\ a2_{3,1} & a2_{3,2} & ... & a2_{3,(q+n)} \\ ... & ... & ... & ... \\ a2_{m,1} & a2_{m,2} & ... & a2_{m,(q+n)} \end{bmatrix}$$

### 3.3 Mage Validation

We now validate the accuracy of Mage's performance inference. Figure 3a shows the estimation error for Mage across 350 application mixes of latency-critical and batch jobs in a simulated 16-core CMP with 4 core configurations, a mix of high-end, and low-power designs (for more details on methodology see Section 5). On average, the error between estimated and measured performance across mixes is 4.6% and at most 9.1%. Figure 3b also shows the estimation error across 160 applications mixes of interactive and batch jobs running on a real cluster with servers of different hardware configurations, with and without power management. The error is again low, 1.7% on average, and up to 5.8% when no power management is used, and 1.9% on average, and up to 6.1%, when DVFS is enabled. SGD is resilient to the specific applications used, with jobs that experience the highest error being mostly volatile interactive workloads that go through drastic load fluctuations at runtime. Low estimation errors ensure that the input Mage uses for scheduling accurately reflects the application's resource requirements, and reduces the need to adjust its decisions frequently once applications are already running.

## 4 RUNTIME

### 4.1 Challenges

Although Mage's initial placement decisions minimize interference, the actual application behavior during runtime may vary for several reasons. First, the application itself

---
SGD2 and SGD3 is lower than the number of iterations it would require to reconstruct randomly-initialized zero columns.

may change characteristics. Most workloads, especially user-driven interactive services go through multiple phases during their execution, each with diverse resource needs. Second, while the techniques used in Mage have high accuracy in the majority of cases, they may still overestimate application performance occasionally. Third, Mage by default optimizes for the mean performance *across* applications on a system. If one or more applications have higher priorities than the rest, this function may penalize their performance. Regardless of the reason behind suboptimal performance, Mage needs to take immediate action to recover from the QoS violation. This creates two challenges: first, the system must be agile in detecting when performance is suboptimal. Interactive applications, such as websearch, must meet strict tail latency requirements, which means that even a few milliseconds of suboptimal performance can have a significant impact on tail request latency [13]. Second, once degradation is detected, Mage should quickly determine whether an alternative placement can resolve the performance issues without incurring disproportionate migration costs.

### 4.2 Fast Detection

Mage runs at the hypervisor level. Inference happens in *Mage Master*, which runs on a dedicated server (see Fig. 1). Each scheduled application runs in a Linux container to isolate its cores from co-scheduled workloads. Threads are additionally pinned to physical cores to avoid interference due to the OS scheduler's decisions [35]. The master spawns a *Mage Agent* on each worker machine that monitors the performance of all scheduled applications, and notifies the master when QoS violations occur. Agents measure low-level performance metrics, such as MIPS and cache misses that applications do not always record. Cloud applications often report their own performance, although this typically reflects high-level metrics, such as request throughput and latency, which the agent can then correlate to low-level statistics.

Apart from measuring application performance, the main goal of the Mage agent is to be unobtrusive and transparent to the application. To ensure this, the performance monitor runs in a separate software thread that wakes up every 1-2 seconds, measures application performance, and goes back to sleep. If the agent detects that the performance of one or more of the scheduled applications deviates significantly from its expected value, it immediately notifies the master over an asynchronous, lightweight RPC protocol [1]. The threshold that signals a QoS violation is configurable; unless otherwise specified we use 10% for the remainder of the paper.

### 4.3 Fast Correction

Once Mage gets notified from one or more agents that application performance is suboptimal it takes action. First, it reprofiles applications online under their current placements. If the measured performance is different from the corresponding column in the utility matrix, Mage replaces the column with the profiling data, and reruns the last (or two last) steps of SGD. There are three possible outcomes from this.

| Servers | Configuration |
|---|---|
| Server1 | 8-core, 1 thread/core, ooo, 2.0GHz, 10 servers |
| Caches | L1: 32KB, private, split D/I, L2: 4MB shared, L3: none |
| Memory | DRAM, 16GB |
| Server2 | 24-core, 2 threads/core, ooo, 2.30GHz, 10 servers |
| Caches | L1: 32KB, private, split D/I, L2: 256KB, private, L3: 16MB, shared |
| Memory | DRAM, 64GB |
| Server3 | 4-core, 4 threads/core, ooo, 3.10GHz, 8 servers |
| Caches | L1: 32KB, private, split D/I, L2: 256KB, private, L3: 8MB, shared |
| Memory | DRAM, 32GB |
| Server4 | 4-core, 2 threads/core, ooo, 1.80GHz, 12 servers |
| Caches | L1: 32KB, private, split D/I, L2: 4MB, shared, L3: none |
| Memory | DRAM, 32GB |

Table 1: The heterogeneous 40-server cluster.

| | Configuration |
|---|---|
| ooo1 | Westmere-like OOO 2.4GHz, 4 cores |
| L1 caches | 32 KB, private, 8-way set-associative, split D/I, 1-cycle latency |
| L2 caches | 256 KB, private, 8-way set-associative, inclusive, 6-cycle latency |
| ooo2 | Westmere-like OOO 2.0GHz, 4 cores |
| L1 caches | 32 KB, private, 8-way set-associative, split D/I, 1-cycle latency |
| L2 caches | 128 KB, private, 8-way set-associative, inclusive, 6-cycle latency |
| ooo3 | Atom x5-z8330-like 1.44GHz, x86-64 ISA; 8B-wide ifetch, 4 cores |
| L1 caches | 128 KB, private, 8-way set-associative, split D/I, 2-cycle latency |
| L2 cache | 12 MB, shared, 16-way set-associative, inclusive, 6-cycle latency |
| in-order1 | In-order 1.6GHz, x86-64 ISA; 8B-wide ifetch, single-issue, 4 cores |
| L1 caches | 16 KB, private, 8-way set-associative, split D/I, 2-cycle latency |
| L2 caches | 128 KB, private, 8-way set-associative, inclusive, 6-cycle latency |
| L3 cache | 12MB, shared, non-inclusive, 20-cycle; 16-way, hashed set-assoc |
| Coherence | MESI, 64B lines, no silent drops; sequential consistency |
| Memory | 64GB, 200 cycles latency, 12.8GBps/channel, 2 channels |

Table 2: The simulated 16-core heterogeneous CMP.

First, Mage determines that there is a better placement that involves context switching applications within a heterogeneous platform. In this case rescheduling is immediate to allow performance to start recovering. The instruction and data footprints of our examined applications are large enough that the overheads associated with context switching and private cache warmup are negligible.

Second, Mage determines that there is a better placement that involves migrating the offending application to another, already utilized server. In this case migration to a new server may come at a significant cost. Mage prioritizes the migration of stateless applications over stateful, and only reschedules stateful applications if no alternative exists. The current system does not support live migration, however, this could help alleviate some of the performance penalties from migration. In case an application has to be migrated, Mage packages its state in its container, sends the image over the network, and resumes execution of the container on the new machine.

Third, Mage determines that there is no better application placement given the currently-available system resources. In that case Mage either migrates one or more stateless applications to a new unused machine, paying a penalty in efficiency, or, if there are low-priority or best-effort workloads, it constrains their resources, and, if needed, terminates them to improve the performance of the high-priority application. In practice, migration across servers is rare, and in most cases it is constrained to stateless workloads (see Sec. 6.2).

Finally, because Mage has global visibility in the cluster state, it can trade off intra-server for inter-server heterogeneity. For example, if a stateless application violates its QoS and scaling up its allocation would cause its co-scheduled applications to suffer, Mage may prioritize migrating the service to fewer resources of a higher-end server, if the resulting performance counterbalances the overhead of migration.

## 5 METHODOLOGY

We evaluate Mage under three scenarios: first, a simulated heterogeneous CMP to show fine-grain scheduling across cores. Second, a heterogeneous physical cluster with different server configurations to show distributed scheduling across machines. Third, using the same heterogeneous cluster we introduce core-level heterogeneity with RAPL and `acpi-cpufreq` to show the hierarchical operation of Mage at inter- and intra-server granularity. Below we describe the systems and applications used in the three scenarios.

**Simulated systems:** We simulate a heterogeneous CMP using *zsim* [50], a fast and scalable multicore simulator. ZSim supports time virtualization to run real-time, interactive applications. Unless otherwise specified, we simulate a 16-core system with four core configurations (4 cores per configuration). Table 2 shows the configuration details.

**Cluster:** We also use a real 40-server cluster with 4 server configurations. The servers vary in terms of their core number and frequency, memory capacity, and storage. They are all connected to 10Gbe links and within 1 network hop from each other. Table 1 provides more details on each platform.

We additionally use RAPL (using the `acpi-cpufreq` driver) to introduce per-core heterogeneity, and to demonstrate Mage's tiered scheduling approach. Mage's inference overheads remain in the scale of 250msec for up to 20 frequency levels per server (uniformly distributed from 1GHz up to the nominal frequency for each server with turbo mode disabled). Cores with frequencies below 1GHz cause tail latency to increase beyond the applications' QoS constraints even running in isolation at low load. In this case Mage first determines the right server platform, given an application's sensitivity across all resources, and then appropriate core frequencies. We assume that during initial scheduling there is no adjustment in the core frequency of any applications already scheduled on a server. Adjustments can happen in subsequent intervals, if they improve the geometric mean of performance across the co-scheduled applications.

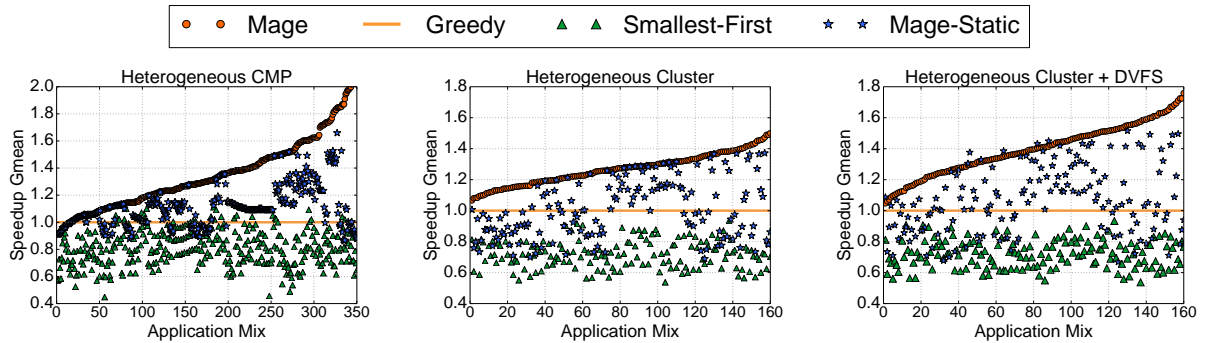

Figure 4: Performance comparison between Mage, Mage-Static, a greedy, and a power-efficient (smallest-first) scheduler for (a) the heterogeneous CMP, (b) the heterogeneous cluster, (c) and the heterogeneous cluster with DVFS.

**Workloads:** We use both latency-critical, interactive services and batch, throughput-bound workloads. In terms of interactive applications, we use *memcached* [26] and *nginx* [3]. *memcached* is an in-memory distributed caching service that is compute and memory-bound. Its $99^{th}$ percentile latency constraint is set at 200usec, consistent with what many cloud operators use [4]. *nginx* is a mostly stateless webserver that works as a front-end for many popular online multi-tier services, and it is primarily compute bound. Unlike memcached, the tail latency constraints for nginx are more relaxed; 10msec for the $99^{th}$%ile, again consistent with reports from cloud operators [10, 12]. The input load for both services is driven by open-loop generators, and follows uniform, exponential, and power-law distributions [35].

In terms of batch workloads, we use the entire SPECCPU2006 suite, and workloads from PARSEC [7], SPLASH-2 [54], BioParallel [29], and Minebench [43]. The ratio of latency-critical to batch applications is 40:60.

## 6 EVALUATION

We first compare Mage to existing scheduling approaches for heterogeneous systems, and then analyze its behavior, overheads, and parameter sensitivity.

### 6.1 Scheduler Comparison

**Performance comparison:** We compare Mage against five schedulers. First, a *Greedy* scheduler that prioritizes scheduling applications to the fastest available core, to optimize performance. This is a common scheduling approach, especially in underutilized systems [5,18,36]. Second, we compare Mage against a power-efficient scheduler (*Smallest-first*), that tries to minimize energy consumption, by first mapping applications on the most energy-efficient cores. Third, against a static version of Mage, *Mage-Static*, where decisions are only made once at the beginning of a program's execution, and not revisited thereafter. Fourth, against PIE [52], a heterogeneity-aware CMP scheduler that uses microarchitectural metrics to determine appropriate application-to-core mappings at runtime. Finally, we compare Mage against Paragon [15], a heterogeneity- and interference-aware cluster scheduler that uses classification to map applications to server platforms, but does not consider intra-server heterogeneity.

Figure 4 shows the comparison between Mage, the Greedy scheduler, and the power-efficient scheduler. Figure 4a shows the results for the simulated heterogeneous CMP, Figure 4b for the heterogeneous cluster without power management and Figure 4c with DVFS. Performance is averaged (*gmean*) across the applications of each mix, and normalized to the performance achieved by Greedy. Mixes are ordered from worst- to best-performing for Mage. For the majority of applications Mage outperforms both Greedy and the power-efficient scheduler. With respect to Greedy, the benefit for Mage comes from using the highest-performant cores only for applications that need them, maintaining high-end resources available for workloads that arrive later. In contrast, Greedy prioritizes the allocation of high-end resources, leaving subsequent applications with suboptimal options. There is a small number of applications for which Greedy outperforms Mage (leftmost part of the graph). These correspond to mixes for which greedy allocation exactly matches the applications' resource requirements. In that case, Greedy avoids the overheads incurred by SGD, as well as any overhead from reprofiling and rescheduling applications at runtime.

The power-efficient scheduler achieves almost always worse performance than Mage, with the exception of a small number of mixes where performance is comparable. This is primarily because *Smallest-first* ignores the resource requirements of incoming applications, and by prioritizing allocation of low-power resources, it exacerbates contention in shared resources. For example, a memory-intensive application running on a low-end core with a small and shallow cache hierarchy, will introduce higher interference in the shared LLC and memory

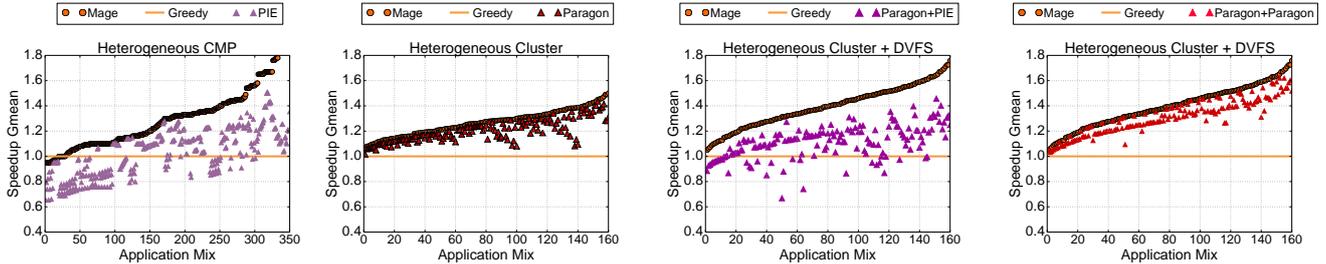

Figure 5: Performance comparison between: (a) Mage and PIE in the heterogeneous CMP, (b) Mage and Paragon in the heterogeneous cluster, and (c) Mage and Paragon+PIE and (d) Mage and Paragon+Paragon in the cluster with DVFS.

system, than if the same application was running in a core with larger private caches.

These results are consistent in the heterogeneous cluster as well (Fig. 4b). Here the deviation between schedulers is even more pronounced, as an incorrect placement cannot easily be corrected by context-switching to another core on the same machine. The performance difference is also high because servers are very diverse, ranging from first generation Nehalem with 1 thread per core and no L3 cache, to the latest Intel Broadwell, with 4 threads per core and 3.10GHz frequency. Introducing power management (Fig. 4c) further penalizes schedulers that ignore per-application resource requirements, as it either maximizes frequency for applications that do not benefit from it (Greedy), or minimizes frequency, hurting performance (Smallest-First).

Figure 4 also compares Mage and *Mage-Static*. Mage-Static behaves considerably better than the power-efficient scheduler, especially when scheduling applications across servers, and in several cases matches Mage in performance. This happens for mixes that do not contend in shared resources, for applications whose resource requirements remain constant throughout their duration, and for applications for which the initial scheduling was correct. There are, however, several mixes for which Mage outperforms the static scheduler. This primarily corresponds to interactive services that go through diurnal patterns. Since the penalty of re-scheduling is greater in a distributed system, Mage-Static is not hurt as much from not revisiting its decisions in Fig. 4b, as for the heterogeneous CMP in which case, if a more performant schedule exists it almost always is adopted. On average Mage outperforms Mage-Static by 22% in the heterogeneous CMP, by 19% in the cluster without DVFS, and by 29% with DVFS.

Finally, Figure 5 compares Mage with recent work on CMP and cluster-level scheduling. In the case of the heterogeneous CMP we compare Mage against PIE [52] (Fig. 5a), in the case of the heterogeneous cluster against Paragon [15] (Fig. 5b), and in the case of the heterogeneous cluster with DVFS against the combination of Paragon and PIE (Fig.5c), and Paragon and Paragon (Fig. 5d), for inter- and intra-server decisions respectively. In this case the two schedulers decide independently, Paragon first selecting the right server for an application, and PIE (or Paragon) then selecting the right core frequency.

Mage significantly outperforms PIE in Fig. 5a, 33% on average, since it can handle more diverse heterogeneous resources, while PIE is geared towards a big versus small core platform, and since Mage additionally accounts for resource contention, while PIE only focuses on the impact of core heterogeneity on performance. In the case of the heterogeneous cluster (Fig. 5b) the difference is small, 8% on average, since Paragon is designed to handle heterogeneity at server granularity. Most of the difference comes from Paragon missing some inter-dependencies between heterogeneity and interference because it is solving the two problems separately for the initial placement, and relies on a greedy scheduler to correct misestimations at runtime. Conversely, Mage leverages a single exploration approach to capture both heterogeneity and interference at once.

Finally, where the schedulers deviate significantly is when introducing core heterogeneity in the cluster (Fig. 5c,d). Despite having heterogeneity-aware schedulers at each level the fact that there is no information exchange between the inter- and intra-server schedulers, hurts performance, by 21% on average for Paragon+PIE, and by 11% for Paragon+Paragon. In contrast, Mage maintains a global view of resource availability and per-application resource requirements, which allows it to account for the trade-offs between faster cores but slower overall servers when making placement decisions. Apart from achieving higher performance, Mage also improves the scalability of inference compared to Paragon, since it uses staged and parallel SGD to obtain the per-application resource requirements. It is also able to handle more diverse heterogeneity, especially at the CMP level, without taking a hit in scalability compared to prior solutions which mostly focus on placing applications on big versus small cores.

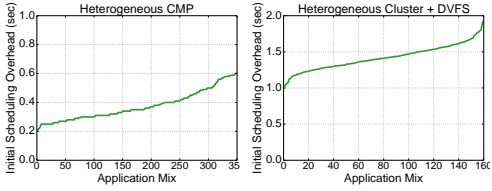
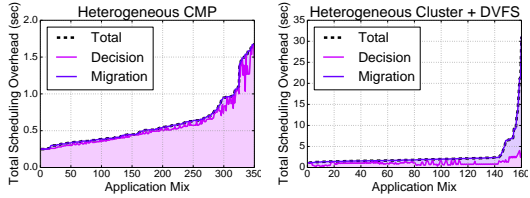
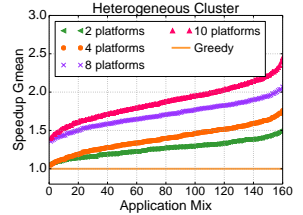

Figure 6: Overhead of initial scheduling for (a) the simulated heterogeneous CMP and (b) the heterogeneous cluster with DVFS.

Figure 7: Scheduling and migration overheads with Mage for the heterogeneous (a) CMP, and (b) cluster with DVFS.

Figure 8: Performance with Mage as server heterogeneity increases.

## 6.2 Mage Analysis

**Scheduling overheads:** Figure 6 shows the overhead Mage adds to application runtime during the initial scheduling phase. Fig. 6a shows the results for the heterogeneous CMP, and Fig. 6b for the heterogeneous cluster with power management; the results are similar for the case without power management. On average overheads are marginal, 0.36sec for the heterogeneous CMP, and 1.28sec for the heterogeneous cluster. The overhead is slightly higher for the cluster because of the added time required to instantiate and initialize containers on different machines. In general, the low scheduling latency ensures that there is no substantial application backlog at admission control. The majority of the overheads in Figure 6a are due to the scheduling algorithm, 76% of total overhead on average, while in Figure 6b the overhead is almost equally distributed between decision and container setup time.

Figure 7a shows the total overheads due to scheduling, throughout the lifetime of scheduled applications for the heterogeneous CMP, and Figure 6b for the heterogeneous cluster with power management. This includes any time Mage detected suboptimal performance for one or more applications in a mix, reprofiled them, ran the staged SGD, and potentially migrated them. For each application mix we report the arithmetic mean across workloads in the mix, and decompose the overhead to scheduling decisions and migration.

In the case of the heterogeneous CMP, most mixes experience very low scheduling delay, 0.52sec on average, and 0.98sec for the $90^{th}$ percentile delay. Almost all the delay comes from the scheduling algorithm, since migration only involves pausing the application on one core, context switching, and resuming execution. In total, Mage reran SGD for 46% of all application mixes, and for 33% of all mixes it rescheduled one or more applications in the mix. The majority of these migrations were caused by changes in application behavior that made the current placement suboptimal. 11% of all mixes were rescheduled more than once, with five mixes experiencing the maximum number of scheduling rounds, *six*, and the highest number of migration rounds, *four*. These mixes correspond to the rightmost part of Figure 7a, and despite the increased scheduling latencies, outperform both the greedy and power-efficient schedulers (see Section 6.1).

The cluster in Figure 7b experiences a different bottleneck. The majority of application mixes experience low overheads, 2.3sec on average and 2.6sec for the $90^{th}$ percentile. Only the rightmost 10 mixes experience higher scheduling overheads than 5sec throughout their execution, and for most of them this amounts to only a small fraction of their total execution time (less than 10% on average). In contrast to the results for the heterogeneous CMP though, here the scheduling algorithm only accounts for a small fraction of the total overheads. The majority of scheduling time now comes from migrating underperforming applications. Migration requires packaging a container and sending it over the network to a different server. To avoid needless migration, Mage prioritizes migrating stateless applications over stateful workloads, such as databases. The latter would suffer from long migration latencies, require substantial time to reinitialize and warm up, and could create network bottlenecks for other nominally operating applications during their migration.

**Sensitivity to the degree of heterogeneity:** We now evaluate how the benefits from Mage scale as we change the degree of heterogeneity in the system. Figure 8 shows the performance gains with Mage as we change the number of different server platforms available in the heterogeneous cluster. The default configuration has 4 server configurations, as discussed in Table 1. A degree of 2 corresponds on one high-end and one low-end server platform, Server2 and Server4 from Table 1 respectively. Similarly, we increase the degree of heterogeneity in the cluster by introducing an additional 4 and 6 server platforms, for a total of 8 and 10 server platforms respectively. The additional platforms range from high-end two-socket platforms, including E5-2699 v4 and E5-2660 v3, to low-power designs, such as the Cavium ThunderX (CN88XX_NT). The total size of the cluster remains the same as before, 40 servers. Finally, for each cluster configuration, we use the same 160 mixes we previously used for all cluster experiments.

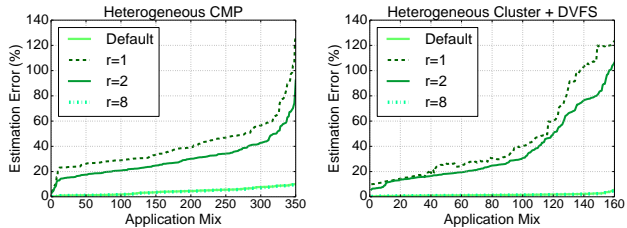

Figure 9: Estimation errors as we vary the density of the input utility matrix for (a) the heterogeneous CMP, and (b) the heterogeneous cluster with DVFS.

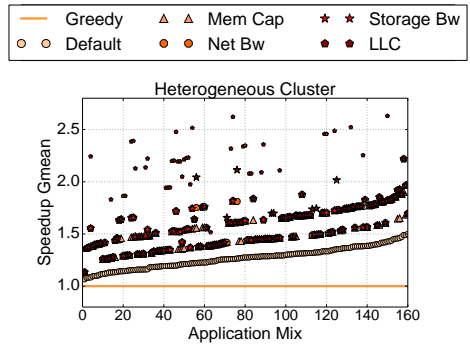

Figure 10: Performance benefits from Mage as we incorporate resource isolation in the heterogeneous cluster.

Figure 8 shows that as the number of heterogeneous server platforms increases, the benefits from using Mage also increase. For example, when there are only two types of servers in the cluster, the performance obtained with Mage is 15% higher compared to the greedy scheduler on average. In comparison, in the cluster with 10 platform configurations, the performance benefit of Mage jumps to 75% on average and up to 2.5x. This is because, intuitively, the more heterogeneous a system becomes, the more likely it is for a heterogeneity-agnostic scheduler to incorrectly map applications on heterogeneous servers, impacting both performance and efficiency. These results are consistent for the heterogeneous CMP as well. When only using two types of cores, Mage achieves 16% better performance on average over the Greedy scheduler, while with 8 types of high-end and low-power cores, the performance improvement is 83% on average, and up to 2.8x.

**Sensitivity to utility matrix sparsity:** By default, Mage uses three training runs, one in isolation, and two with two contentious kernels of tunable intensity. Figure 9 shows how the performance estimation error changes as we vary the density of the utility matrix during training. When Mage only uses one profiling run of the application running in isolation, the average error is 45% for the CMP, and 41% for the cluster when DVFS is used (the results are similar without power management). When using one run in isolation and a single run with a contentious kernel the error drops to 29% on average in Figure 9a, and 23% in Figure 9b. However, given the strict QoS constraints that cloud applications must meet, this error is still unacceptable. Adding one more data point with another contentious kernel reduces the error to 4.7% on average for the CMP, and 1.7% on average for the heterogeneous cluster; the default values used in the rest of the paper. Increasing the utility matrix density further does not significantly impact the estimation accuracy.

**Resource isolation:** Mage already uses containers and pins threads to physical cores, to eliminate interference from the OS scheduler [35]. It also uses DVFS to determine whether frequency scaling is beneficial. In the remaining resources, it leverages heterogeneity to reduce shared resource contention, but cannot entirely eliminate it. Recently, more isolation techniques have been integrated in modern platforms, including memory and cache capacity partitioning, and network and storage bandwidth [39] isolation. However, resources such as memory bandwidth, TLBs, and private caches (L1, L2) can still not be isolated, allowing some interference to endure. We now expand Mage to consider partitions in other resources apart from cores and the power budget (through DVFS). We progressively introduce memory capacity and storage bandwidth partitioning through cgroups, network bandwidth partitioning using the Linux traffic scheduler (LTS) [39], and cache partitioning using Intel's Cache Allocation Technology (CAT) [2]. We consider memory capacity partitions at the granularity of 4GB, network and storage bandwidth at the granularity of 10% of the maximum provisioned bandwidth, and LLC capacity at the granularity of 2 ways. This increases the number of columns in SGD by 2-3 orders of magnitude depending on the machine, and the corresponding inference overheads by 1-2 orders of magnitude, amounting to 2.3sec on average. Although the increase is significant, it also translates to higher speedups compared to the default scheduler. Because of the staged, parallel SGD, the increased problem size does not also correspond to higher scheduling overheads.

Figure 10 shows how performance with Mage changes as we incorporate isolation in the heterogeneous cluster. Partitioning memory capacity substantially improves performance, with the average speedup over Greedy being 51% and up to 90.1%. Network bandwidth helps network intensive applications like memcached, but does not impact the rest of the workloads. Similarly, storage bandwidth isolation does not have a major impact on performance, since none of the examined workloads really stress persistent storage. Finally, partitioning the last level cache has the largest impact on performance for applications whose working sets fit in the cache, and which previously suffered from being co-scheduled with cache thrashing workloads. In general, isolation complements heterogeneous scheduling by allowing it to reach the full potential of the heterogeneous resources. One can reduce the overheads of sizing resource partitions by using Mage to obtain coarse-grained insights on the benefit of different resources, and fine-tuning allocations at runtime.

# 7 CONCLUSIONS

We have presented Mage, a practical, online runtime that manages intra- and inter-server heterogeneity, and minimizes resource contention in cloud systems. Mage leverages a set of scalable machine learning techniques, including stochastic gradient descent to quickly determine the joint impact of heterogeneity and interference on application performance. We have validated the accuracy of Mage's performance estimations, and have evaluated the runtime both using simulations and real cluster experiments. In all cases Mage improves performance compared to greedy, power efficient, and previous heterogeneity-aware CMP and cluster schedulers, while also improving the latency of scheduling decisions.